\documentclass[prc,showpacs]{revtex4}

\begin{document}

\title{CAN DEUTERIUM FORMATION BE MEASURED?} 

\author{Robert K.\ Soberman}
\email{soberman@sas.upenn.edu}
\homepage{http://www.sas.upenn.edu/~soberman}
\altaffiliation{2056 Appletree Street, Philadelphia, PA 19103}
\author{Maurice Dubin}
\email{mdubin@aol.com}
\altaffiliation{14720 Silverstone Drive, Silver Spring, MD 20905}
\affiliation{retired}

\date{\today}
 
\begin{abstract}  

The fundamental PP reaction has never been attempted in the laboratory as theory states it cannot be measured within the human life span. This forms the foundation of stellar nucleosynthesis and the standard solar model.  Yet numerous observations including the Sun's obvious variability argue the theory is flawed and the reaction occurs in times many orders of magnitude shorter than prediction. Considering the consequence to accepted models and controlled fusion, testing is warranted. 

\end{abstract}


\maketitle 

There is no record of an attempt to measure the transmutation probability for the fundamental reaction $\rm ^1H(p,e^+,\nu)^2H$.  As Bethe and Critchfield \cite[]{Bet38} predicted a reaction rate of $14(10)^9$ years no one has been bold enough to test this prediction.  Parker and Rolfs \cite[]{Par91} wrote ``It can be estimated that with a total cross section of $\rm 10^{-47} cm^2$ at $\rm E_p(lab) = 1 MeV$ for a proton beam of 1 mA incident on a thick hydrogen target, there would be only one $\rm ^1H(p,e^+,\nu)^2H$ reaction in $10^6$ yr.'' A similar but numerically discrepant statement appears in Clayton's  classic stellar nucleosynthesis text \cite[]{Cla68}.  As no research team has the patients, resources or longevity to wait out such a prediction it is not surprising that it has never been tested. 

Such a slow reaction rate was necessary for the standard stellar power model.  A much shorter time would cut the lifetime of stars accordingly and a brief reaction time such as seconds would, in that model, cause stars to explode when the reaction conditions were reached. Yet the theorized extremely small hydrogen transmutation probability leads to derivative conclusions that are inconsistent with measurement.  Foremost is the deuterium to hydrogen ratio.  As the 
$\rm ^2H(p,\gamma)^3He$ reaction has high probability, deuterium is believed to be destroyed early in stellar formation \cite[]{Eze66}.  The model predicts a deuterium to hydrogen ratio of $10^{-17}$ in stellar cores \cite[]{Bah92} twelve orders of magnitude less than the $10^{-4}$ to $10^{-5}$ ratios observed in nature.  There exist a number of other indications (below) that deuterium forms much more easily than predicted.

When Bethe and Critchfield \cite[]{Bet38} proposed hydrogen fusion as the long lasting energy source capable of sustaining the Sun and stars over Gigayears, it was understandable that they looked to the core for the requisite temperature ($\rm >\,10^7 K$) and pressure ($\rm >\,10^5 kg/m^3$). Space age technology was needed to discover that fusion could occur at conditions existing near the surface. Terekov et al.\ 
\cite[]{Ter93} reported observing the 2.2 MeV gamma ray line of the 
$\rm ^1H(n,\gamma)^2H$ reaction in the solar flare of May 24, 1990.

A sizeable quantity of radioactive $\rm ^7Be$ (53-day half-life) was unexpectedly found on the leading surfaces of the Long Duration Exposure Facility (LDEF) that had orbited the Earth for 69 months at an altitude of 310 km \cite[]{Fis91,Phi92}.  Cosmic ray spallation was ruled out as the production source \cite[]{Gre93} after examining the LDEF surfaces for similar traces of $\rm ^{10}Be$ ($1.5(10)^6$ yr half-life). $\rm ^7Be$ is a product of a lesser probability branch of the PP reaction chain (PPII).  To explain the $\rm ^7Be$ discovery Share and Murphy \cite[]{Sha97} proposed that it was produced during the intense solar flares that occurred in late 1989.  A problem with their explanation is the 100\% Earth capture efficiency they required to fit the measurement.  Chappell et al.\ \cite[]{Cha87} showed that only about $10^{-3}$ of the solar wind penetrates the terrestrial bow shock.  This must also apply to the ionized $\rm ^7Be$.  The absence of $\rm ^7Be$ spectra measurements evidences its ionized state.  Thus the flare produced deuterium production scenario fails by several orders of magnitude to produce the LDEF measured quantity of $\rm ^7Be$.  No other source of sufficient strength exists within a travel time of several half-lives.  The international space station provides an ideal laboratory for $\rm ^7Be$ study.  Extra-vehicular activities are frequently required.  Material samples may be exposed on a leading surface for several months during periods of reduced solar activity, or direct measurement of the structure's surface would confirm a continuous $\rm ^7Be$ background. 

If the $\rm ^7Be$ collected 310 km above the Earth was not produced by cosmic rays or in solar flares, whence its source?  The 53-day half-life excludes extra-solar system or solar core production.  A proposed solution provides the LDEF measured quantity of $\rm ^7Be$ \cite[]{Sob01}.  The source is near solar surface proton-proton fusion.  Weak hydrogen dominated interstellar meteoroids \cite[]{Dub91} arrive at the surface with near solar escape velocity (618 km/s; $15(10)^6$ K thermal equivalent). Several have observed and measured this hyper velocity solar directed flux \cite[]{Hic74,Fri78,Bea80}.  The structurally weak partly ionized aggregates are drawn into gravity driven plasma vortices that produce a high density pinch.  Still cold from space, molecular water spectra observed in 4,000 K sunspots \cite[]{Woh69} evidence this inflow.  Interacting with the solar atmosphere at high velocity and high pressure, hydrogen is transformed to helium with appropriate byproducts such as $\rm ^7Be$.  However, if the reaction were as improbable as Bethe and Critchfield \cite[]{Bet38} theorized it could not provide the LDEF result \cite[]{Fis91,Phi92}.

Apart from providing a solution to the long standing deuterium to hydrogen ratio dilemma, near prompt stellar surface PP fusion answers numerous observational enigmas.  One of these is the "the lithium problem" \cite[]{Boh92}.  $\rm ^7Li$ is observed in the Fraunhofer spectra of our Sun and atmospheric spectra of many main sequence stars.  It is generally ascribed to cosmic ray spallation.  However, the relative absence of another stable isotope, $\rm ^6Li$, that should also be created by spallation poses a serious problem for that explanation. It is easily explained as the daughter of $\rm ^7Be$ decay but the short $\rm ^7Be$ half-life excludes core-surface transport. Deuterium abundance and the lithium problem represent but two examples of discrepant isotopic abundance. Complex theories \cite[]{Fow62} are required absenting near prompt stellar surface PP fusion to account for measured low mass isotopes. $\rm ^7Li$ is also common in nova spectra on the surface of white dwarf stars \cite[]{Sta78}. The $\rm ^7Be$ gamma ray line is also observed there \cite[]{Lei90}.  Inventive theories such as slow precursor PP fusion \cite[]{Sta89} are created to account for their presence. 

To explain PP fusion produced elements observed in population II red giants and associated nebulae, a process called ``dredge-up'' is hypothesized.  It contends that circulation extending from the core to the surface raises fusion products to be ejected in the stellar wind.  How hydrogen is retained in the core in the presence of such circulation is ignored. 

Optical opacity of the thin ($\rm \sim\,100 km$) photosphere likely prevents direct view of near surface solar PP fusion although some may be associated with chromospheric UV ``explosive events.''  Larger reactions may be what are referred to as ``bright points'' \cite[]{Gol74,How79}, sporadic transient isolated x-ray flashes with temperatures exceeding $10^6$ K that are regularly observed.  About 1500 are estimated on the Sun at any given moment but they remain enigmatic.  

Another astronomical phenomenon that signals a much less than predicted PP reaction time is short period stellar luminosity variation.  A photon takes about $10^7$ years to random scatter from the core to the surface. Physics does not permit cyclic variation with shorter periods to persist through such transit. To explain variations with periods that range from hours to years, surface opacity change often requiring sub-surface luminosity variation is hypothesized. The cause, if addressed, is problematic.

Flares, mass ejection and other forms of surface activity require huge quantities of local energy.  Magnetic reconnection, a phenomenon that has never been demonstrated in the laboratory nor shown theoretically to work in the proper time is accepted for lack of alternatives.  Sporadic prompt near surface PP fusion initiated by the impact of plasmas generated from the arrival of interstellar meteoroids provides the necessary energy.  It also explains the detection of 0.511~MeV gamma rays that have been observed in flares \cite[]{Ram79}.  This gamma ray line is the consequence of electron positron annihilation, the positrons originating from deuterium formation.  There is too a sizeable unexplained 0.511~MeV sky background. Large excesses of 
$\rm ^3He$, a product of PP chain fusion are also associated with solar flares \cite[]{Sch62}. 
 
Neutrinos from the sun provide the strongest evidence of fault in the standard solar model (SSM).  From the earliest measurements \cite[]{Dav68} they have been significantly different from prediction \cite[]{Bah89}.  Neutrino ``telescopes'' around the world have been in part justified to investigate the solar neutrino ``shortfall.'' This has become a fundamental nuclear physics question as the proffered solution of neutrino flavor transition requires a yet to be detected neutrino mass.

Solar power varies appreciably over time including an approximate 11-year cycle.  Yet observers refuse to believe their senses and instruments because theory says otherwise.  Early measurements from the Sudbury Neutrino Observatory (SNO) yielded a high solar neutrino count \cite[]{Ahm01} compared to Super Kamiokande \cite[]{Fuk01}.  Theorists point out that the ``formerly missing electron neutrinos'' had undergone the predicted flavor change to mu and tau neutrinos and are now being tabulated \cite[]{Bah01}.  Ignored is the temporal difference between SNO measurements taken near the peak of the solar cycle and the longer operating Super Kamiokande cycle average because the SSM denies variability in times less than millions of years. When a similar difference ``existed'' between the Homestake neutrino observatory and Kamiokande, Davis \cite[]{Dav94} showed the difference disappeared when near contemporaneous measurements from the two detectors were compared.  His paper was criticized because he used the SSM to normalize results from the two (to avoid repeating tedious calculations) but discounted the SSM prediction. Unstated was the objection to Davis' implication that the solar fusion rate varied through the solar cycle. Obvious variation in the data of the several neutrino experiments through the solar cycle has been consistently attributed to statistics. The SSM is silent on the physically disallowed maintenance of an approximate 11-year cycle of surface activity powered by a millennial invariant core.

Statistics were stretched almost beyond the elastic limit to discount Homestake run number 117 that extended over several solar flares including the largest ever recorded. That run yielded a neutrino count about five times those preceding and following and six times the long term mean \cite[]{Dav94}.  Fairness demands noting that the 19 
$\rm ^{37}Ar$ atoms counted from run 117 were only 3 sigma above the mean.  While extremely low counts ($\rm \sim 0.5\,day^{-1}$) render the only occasion for this large departure from the norm statistically expected, the coincidence with extreme solar activity takes it from improbable to neigh impossible if fusion power is invariant.   

If PP deuterium fusion does occur quickly as the foregoing would indicate, what might be the mean reaction time compared to the 
$\rm 14(10)^9 yr$ theorized? An examination of the relation for the mean reaction time $t$ provides a clue.  

$$t \; = \; {\rho_i \over {r(\sigma \rho_i \rho_d T)}}$$

The mean reaction time is inversely proportionate to the rate $r$ with which reactions occur.  That rate depends upon the cross section $\sigma$, the temperature $T$, the densities of interacting nuclei 
$\rho_i$ and varies near linearly with the resultant daughter nuclei $\rho_d$. The theoretical slow mean time of $14(10)^9$ yr is consistent with hypothesized stellar life but results in a $10^{-17}$ deuterium/hydrogen ratio in star cores.  This value is at least twelve orders of magnitude less than measured anywhere including stellar atmospheres, stellar winds and the so-called ``dredge-ups'' discussed above. Supposed core fusion destroys deuterium as soon as it is formed so its presence in quantity poses a problem to the generally accepted model \cite[]{Sch65}. However, if the observed $\rm ^2H/^1H$ ratio of $10^{-4}$ to $10^{-5}$ is the result of PP reaction then the relation above would suggest that the hypothetical reaction rate and mean reaction time are in error by 12 to 13 orders of magnitude or more. 

In view of the import of the fundamental PP reaction, any question of its parameters need be examined. Consider the consequence resultant to a measurement differing from prediction.  No need to wait $10^{10}$ years.  The absence of evidence for the reaction, i.e. the 0.511~MeV gamma ray, for a period of hours would quash the concept of near stellar surface hydrogen fusion \cite[]{Sob01} and any prompt fusion suggestions.  It is possible that conditions present in stellar fusion, such as the presence of small amounts of other elements (e.g. C, N, O) may be required to imitate nature. 

The data may already exist.  Established theory is a strong deterrent to publishing contrary observations.  Stalwart researchers and sympathetic journal referees are necessary for such measurements to 
draw attention. 

One test, itself in question, is hydrogen fusion during acoustic cavitation.  Taleyarkhan et al.\ \cite[]{Tal02} published evidence for DD fusion during cavitation in deuterium loaded acetone bombarded by energetic neutrons. Excess tritium and energetic neutrons were reported; absent in control experiments with normal acetone.  It is likely that 0.511~MeV gamma rays, {\it if observed}, would have been ignored as PP fusion, {\it if it occurred}, would only compound the controversy. As this experiment will likely be duplicated by those authors and others, examination for positron electron annihilation would be a relatively simple addition.


\begin{thebibliography}{}

\bibitem[1]{Bet38} H. A. Bethe and C. L. Critchfield, Phys. Rev. 
{\bf 54}, 248 (1938).

\bibitem[2]{Par91} P. D. M. Parker and C. E. Rolfs, in {\it Solar Interior and Atmosphere}, edited by A. N. Cox, W. C. Livingston and   M. S. Matthews (University of Arizona Press, Tucson, 1991) p. 31.

\bibitem[3]{Cla68} D. D. Clayton, {\it Principles of Stellar Evolution} (McGraw-Hill, New York, 1968).

\bibitem[4]{Eze66} D. Ezer and A. G. W. Cameron, in {\it Stellar Evolution}, edited by R. F. Stein and A. G. W. Cameron (Plenum Press, New York, 1966) p. 203. 

\bibitem[5]{Bah92} J. N. Bahcall and M. H. Pinsonneault, Revs. Mod. Phys. {\bf 64}, 885 (1992).

\bibitem[6]{Ter93} O. V. Terekhov, R. A. Syunyaev, A. V. Kuznetsov,  
C. Barat, R. Talon, G. Trottet and N. Vilmer, Astron. Lett. {\bf 19}, 65 (1993). 

\bibitem[7]{Fis91} G. J. Fishman, {\it et al.},  Nature, {\bf 349}, 678 (1991).

\bibitem[8]{Phi92} G. W. Phillips, S. E. King, R. A. August,  
J. C. Ritter, J. H. Cutchin, P. S. Haskins, J. E. McKisson, 
D. W. Ely, A. G. Weisenberger, R. B. Piercey and T. Dybler,  
in {\it LDEF - 69 Months in Space, Proceedings of the First Post-Retrieval Symposium, {\bf CP-3134}} (NASA, Washington, 1992) 
p. 225. 

\bibitem[9]{Gre93} J. C. Gregory, A. Albrecht, G. Herzog,  J. Klein, 
R. Middleton, B. Dezfouly-Arjomandy and B. A. Harmon,  in {\it LDEF - 69 Months  in Space, Proceedings of the Second Post-Retrieval Symposium, {\bf CP-3194}} (NASA, Washington, 1993) p. 231.

\bibitem[10]{Sha97} G. H. Share and R. J. Murphy, ApJ {\bf 485}, 409 (1997).

\bibitem[11]{Cha87} C. R. Chappell, T. E. Moore and J. H. Waite, Jr.  JGR {\bf 9}, 5896 (1987).

\bibitem[12]{Sob01} R. K. Soberman and M. Dubin, {\it Dark Matter Illuminated} (Infinity Publishing.com, Haverford, 2001) p. 40.

\bibitem[13]{Dub91} M. Dubin and R. K. Soberman, Plan. Space Sci. 
{\bf 39}, 1573 (1991). 

\bibitem[14]{Hic74} T. R. Hicks, B. H. May and N. K. Reay, MNRAS 
{\bf 166}, 439 (1974).

\bibitem[15]{Fri78} J. W. Fried, A\&A {\bf 68}, 259 (1978).

\bibitem[16]{Bea80} W. I. Beavers, J. J. Eitter, P. H. Carr and  
B. C. Cook, ApJ {\bf 238}, 349 (1980). 

\bibitem[17]{Woh69} H. W\"ohl, Solar Phys. {\bf 9}, 394 (1969).

\bibitem[18]{Boh92} E. B\"ohm-Vitense, in {\it Introduction to Stellar Astrophysics, Vol. {\bf 3} Stellar Structure and Evolution} (Cambridge U., Cambridge, 1992) p. 84.

\bibitem[19]{Fow62} W. A. Fowler, J. L. Greenstein and F. Hoyle,   Geophys. J. {\bf 6}, 148 (1962).

\bibitem[20]{Sta78} S. Starrfield, J. W. Truran, W. M. Sparks and  
M. Arnould, ApJ {\bf 222}, 600 (1978).

\bibitem[21]{Lei90} M. D. Leising, in {\it Gamma-Ray Line Astrophysics, AIP Conf. Proc. {\bf 232}} edited by P. Durouchoux and N. Prantzos  (AIP, New York, 1990) p. 173.

\bibitem[22]{Sta89} S. Starrfield, in {\it Classical Novae}, edited by M. Bode (Wiley, New York, 1989) p. 39. 

\bibitem[23]{Gol74} L. Golub, A. S. Krieger, J. K. Silk,  
A. F. Timothy and G. S. Vaiana, ApJ {\bf 189}, L93 (1974).

\bibitem[24]{How79}  R. Howard, L. Fritzov\'a-\v{S}vestkov\'a and 
Z. \v{S}vestka, Solar Phys. {\bf 63}, 105 (1979).

\bibitem[25]{Ram79} R. Ramaty and R. E. Lingenfelter, Nature {\bf 278}, 
127 (1979).

\bibitem[26]{Sch62} O. A. Schaeffer and J. Zahringer, Phys. Rev. Lett. {\bf 8}, 389 (1962).

\bibitem[27]{Dav68} R. Davis, Jr., D. S. Harmer and K. C. Hoffman,  Phys. Rev. Lett. {\bf 20}, 1205 (1968). 

\bibitem[28]{Bah89} J. N. Bahcall, {\it Neutrino Astrophysics} (Cambridge U. Press, Cambridge, 1989). 

\bibitem[29]{Ahm01} Q. R. Ahmad, {\it et al.}, Phys. Rev. Lett. 
{\bf 87}, 071301 (2001).

\bibitem[30]{Fuk01} S. Fukuda, {\it et al.}, Phys. Rev. Lett. {\bf 86}, 5656 (2001).

\bibitem[31]{Bah01} J. N. Bahcall, M. H. Pinsonneault and S. Basu, 
ApJ {\bf 555}, 990 (2001).

\bibitem[32]{Dav94} R. Davis, Jr., Prog. in Part. and Nucl. Phys. 
{\bf 32}, 13 (1994).

\bibitem[33]{Sch65} M. Schwarzschild, {\it Structure and Evolution of the Stars} reprint of (Princeton University Press, Princeton, 1958) (Dover, New York, 1965).

\bibitem[34]{Tal02} R. P. Taleyarkhan, C. D. West, J. S. Cho,  
R. T. Lahey, Jr., R. I. Nigmatulin and R. C. Block, Science {\bf 295}, 1868 (2002).

\end{thebibliography}
\end{document}